\documentclass[runinaddress,showpacs,twocolumn,prb]{revtex4}
\usepackage{amsmath}
\usepackage{amssymb}
\usepackage{graphicx}
\usepackage{bm}

\begin{document}

\title{Disorder Effects in the Quantum Hall Effect of Graphene p-n Junctions}
\author{Jian Li and Shun-Qing Shen}
\affiliation{Department of Physics, and Center for Theoretical and Computational Physics,
The University of Hong Kong, Pokfulam Road, Hong Kong, China}
\date{\today}

\begin{abstract}
The quantum Hall effect in graphene p-n junctions is studied numerically
with emphasis on the effect of disorder at the interface of two adjacent
regions. Conductance plateaus are found to be attached to the intensity of
the disorder, and are accompanied by universal conductance fluctuations in
the bipolar regime, which is in good agreement with theoretical predictions
of the random matrix theory on quantum chaotic cavities. The calculated Fano
factors can be used in an experimental identification of the underlying
transport character.
\end{abstract}

\maketitle

\section{Introduction}

When a mono-layer of honeycomb lattice is singled out of graphite,\cite%
{Novoselov04} this two-dimensional material, dubbed graphene, acquires
extraordinary electronic properties.\cite{Geim07, Katsnelson07, Neto07091163}
Electrons in graphene mimic massless Dirac fermions with extremely high
mobility and tunability,\cite{Novoselov05nat,Zhang05nat} which makes this
material interesting both theoretically and practically. The tunability of
the carrier type via the electric-field effect, in particular, allows for
the realization of graphene p-n junctions using only electrostatic gating.%
\cite{Williams07sci, Ozyilmaz07prl} The quantum Hall effect in these
graphene p-n junctions has shown new fractional plateaus in the bipolar
regime\cite{Williams07sci} that were explained by uniform mixing among edge
states at the junction interface.\cite{Abanin07sci} The mechanism of the
mode mixing, however, is still unclear.

In this paper we address this problem by investigating the transport
characteristics of graphene p-n junctions in the quantum Hall regime with
disorder at the junction interface. Our calculations are based on the
Landauer-B\"{u}ttiker formalism for coherent transport,\cite{LB} and the
results are explained using the random matrix theory (RMT) of quantum
transport.\cite{BarangerJalabert94, Beenakker97rmp}

\begin{figure}[tbp]
\centering  \includegraphics[width=0.45\textwidth]{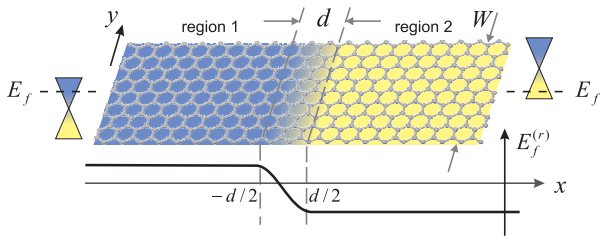}
\caption{(Color online) Schematic diagram of a graphene junction.
Two locally gate-controlled regions of a graphene strip (width
$W$), connected with a
reservoir at each of the far ends, are jointed by a transition area (length $%
d$) where potential is assumed to be composed of a slope along the strip and
a random distribution over each site within. The lower curve shows the
profile of the relative Fermi energy $E_{f}^{(r)}$.}
\label{fig:setup}
\end{figure}

\section{Model and method}

The setup of our simulation is illustrated in Fig. \ref{fig:setup}. A
graphene strip of width $W$ is divided into two regions by a transition area
with length $d$. In either region the carrier type and density can be
locally tuned through an electrostatic gating, and the relative value of the
Fermi energy to the local charge neutrality point is defined as the relative
Fermi energy $E_{f}^{(r)}$. The whole sample is subject to a perpendicular
magnetic field, and the Landau levels are formed in the quantum Hall regime.%
\cite{Gusynin05prl} With a specific Fermi energy, the filling factor of
Landau levels is scarcely dependent on the details of the sample edge,
provided the sample size is big enough,\cite{note1} therefore we use samples
with zigzag edges to carry out our simulation but the results are applicable
to general cases. On the energy scale of our problem, both Zeeman splitting
and spin-orbit interaction are negligibly small, different spin states can
be taken as degenerate and uncorrelated, thus we assume spin is irrelevant
in our calculation and simply multiply the result by a factor accounting for
the spin degree of freedom. For this reason the filling factors $\nu _{1}$
and $\nu _{2}$ given below are all "spinless", i.e., their values are $\pm 1$%
, $\pm 3$, $\pm 5\cdots $ instead of $\pm 2$, $\pm 6$, $\pm 10\cdots $.

Magneto-conductance of a graphene p-n junction has been theoretically
discussed in terms of its valley isospin dependence.\cite{GPNJtheories}
These discussions in general assume an absence of intervalley scattering
across the graphene p-n junction. To explain the fractionally quantized
plateaus observed in the experiments,\cite{Williams07sci, Ozyilmaz07prl}
however, full mode mixing was required in random matrix theory, which
inevitably involves intervalley scattering\cite{Abanin07sci}. We attribute
the inter-valley scattering in our model to disorder potential which varies
quickly enough in the scale of the lattice constant. Disorder in graphene
may have various sources, and the understanding of its role in transport
properties is still incomplete.\cite{Neto07091163, Pereira08prb}

In this paper we focus on the effect of the random potential at the
interface of the junction. The reason is that current inside either region 1
or region 2 is carried by quantum Hall edge states,\cite{Brey06prb} thus
immune to most disorder inside either region, while scattering among states
at the interface of the two regions contributes to the major effect of
disorder on the overall conductance. Disorder at the interface may come from
intrinsic sources like vacancies and impurities, or from extrinsic sources
like random potential introduced by the irregularities of the gate edge. We
model the disorder potential at the interface by using the Anderson-type
on-site energy\cite{Anderson58} which varies randomly from site to site
within the transition area, where a potential slope connecting two sides of
the junction serves as the background potential. Hence the total Hamiltonian
reads
\begin{equation}
H=\sum\limits_{i}\varepsilon (\bm{r}_{i})c_{i}^{\dagger
}c_{i}-\sum\limits_{\langle i,j\rangle }[te^{i\phi _{ij}}c_{i}^{\dagger
}c_{j}+h.c.]  \label{eq:ham}
\end{equation}%
where $c_{i}^{\dagger }$ and $c_{i}$ are the electron creation and
annihilation operators at site $\bm{r}_{i}\equiv (x_{i},y_{i})$,
respectively, $t\approx 2.8$ eV is the nearest neighbor hopping energy in
the graphene lattice and $\langle i,j\rangle $ stands for a
nearest-neighboring pair,
\begin{equation}
\phi _{ij}=\frac{e}{\hbar }\bm{A}(\frac{\bm{r}_{i}+\bm{r}_{j}}{2})\cdot (%
\bm{r}_{i}-\bm{r}_{j})
\end{equation}%
is the phase acquired when an electron hopping from $\bm{r}_{j}$ to $\bm{r}%
_{i}$ in an external field B described by vector potential $\bm{A}$, and the
on-site energy $\varepsilon (\bm{r}_{i})$ are $\varepsilon _{1}$ and $%
\varepsilon _{2}$ in region 1 and 2, respectively;
\begin{equation}
\varepsilon (\bm{r}_{i})=(\varepsilon _{2}-\varepsilon _{1})\frac{x_{i}}{d}+%
\frac{1}{2}(\varepsilon _{1}+\varepsilon _{2})+\mbox{rand}(\Delta )
\end{equation}%
in the transition area ($|x_{i}|\leq d/2$) with rand$(\Delta )$ a random
number uniformly distributed in $[-\Delta /2,\Delta /2]$. The width of the
junction in our simulation is taken to be $W=200(\sqrt{3}a/2)\approx 42.6$
nm, where $a\approx 0.246$ nm is the lattice constant of graphene, and the
length of the interface area is taken to be $d=20a\approx 4.9$ nm. Landau
gauge $\bm{A}(\bm{r})=(-By,0)$ is adopted, with the magnitude of the
magnetic field $B\approx 113$ Tesla which is equivalent to a magnetic flux $%
\Phi =h/701e$ in each unit cell. The magnetic length in this case is $l_{B}=%
\sqrt{\hbar /(eB)}\approx 2.4$ nm, which is about a half of the length of
the interface area, or $1/18$ of the width of the sample. It should be
mentioned here that in a real sample which is presumably much larger in
size, the magnetic field necessary for the quantum Hall effect can be much
smaller.

\begin{figure}[tbp]
\centering \includegraphics[width=0.4\textwidth,
height=7cm]{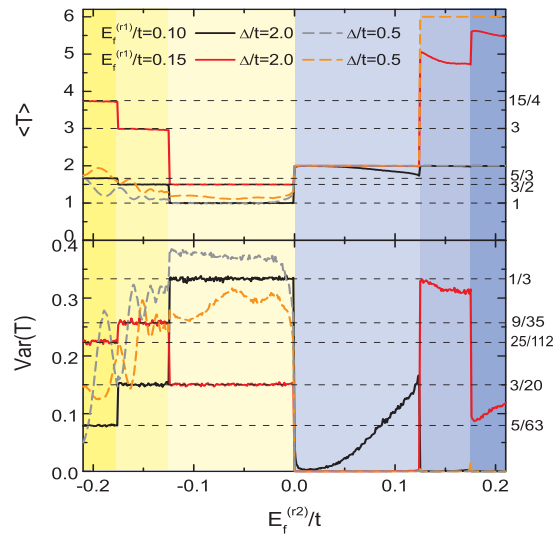} \caption{(Color online) (a) The mean
and (b) the variance of the transmission function $T$ as functions
of the relative Fermi energy $E_f^{(r2)}$ in region 2, with fixed
relative Fermi energy $E_f^{(r1)}$ in region 1 and fixed disorder
strength $\Delta$. Cases with different $E_f^{(r1)}$ and different
$\Delta$
are also compared. $E_f^{(r1)}>0$ implies n-type of region 1, $%
E_f^{(r1)}/t=0.10$ corresponds to the filling factor $\protect\nu_1=1$ and $%
E_f^{(r1)}/t=0.15$ corresponds to $\protect\nu_1=3$.}
\label{fig:diff_Ef_r2}
\end{figure}

We assume coherent transport in the graphene junction, where the Landauer-B%
\"{u}ttiker formalism can be applied.\cite{LB} Transmission functions $T_{pq}
$ ($p,q=1,2$ and $p\neq q$) of the junction are calculated by using the
recursive Green's function technique.\cite{Lee81prl,Li} Vanishing net
current in equilibrium implies that $T_{21}=T_{12}=T$, therefore the
conductance is proportional to either of the transmission functions, $%
G=(e^{2}/{h})T$, where the spin degeneracy has been included in $T$, and the
variance of the conductance $\mbox{Var}(G)=(e^{2}/{h})^{2}\mbox{Var}(T)$,
where $\mbox{Var}(T)$ represents the variance of $T$. In the following we
will be satisfied with observing only the behavior of $T$ in different
situations. Each situation, that is, each experimental condition under which
measurements are made, is identified with a specified combination of $%
\varepsilon _{1}$, $\varepsilon _{2}$ and $\Delta $, while various
configurations of disorder are subject to some self-averaging process in
each measurement. This self-averaging process could be a result of time
dependent electric field used in the experiments,\cite{Abanin07sci} and will
suppress the fluctuation of the measured conductance, thus makes the mean
value a reasonable account for the experimental observation. Our calculation
extracts the mean and the variance of the transmission functions $T$ in each
situation from output of 40,000 samples with different disorder
configurations.

\section{Results}

The calculated transmission functions as shown in Fig. \ref{fig:diff_Ef_r2}%
(a) have surprisingly recovered the quantized transport plateaus observed in
the experiment by Williams et al.\cite{Williams07sci} In junctions with the
disorder strength $\Delta =2t$ (solid lines in Fig. \ref{fig:diff_Ef_r2}),
the ensemble average of the transmission functions form nearly perfect
plateaus in the bipolar regime ($E_{f}^{(r2)}<0$), and the height of each
plateau is
\begin{equation}
\langle T\rangle =2\times \frac{|\nu _{1}\nu _{2}|}{|\nu _{1}|+|\nu _{2}|}
\label{eq:tfnp_mean}
\end{equation}%
with $\nu _{\mbox{\scriptsize total}}\equiv |\nu _{1}|+|\nu _{2}|$.
Corresponding to each plateau of the averaged transmission function, the
ensemble variance of $T$ also develops into a plateau described by
\begin{equation}
\mbox{Var}(T)=4\times \frac{(\nu _{1}\nu _{2})^{2}}{\left( |\nu _{1}|+|\nu
_{2}|\right) ^{2}[(|\nu _{1}|+|\nu _{2}|)^{2}-1]}.  \label{eq:tfnp_var}
\end{equation}%
Both Eq. \eqref{eq:tfnp_mean} and Eq. \eqref{eq:tfnp_var} are the
predictions of the RMT on a quantum chaotic cavity,\cite{Beenakker97rmp}
with the additional factors 2 and 4 from the spin degeneracy. In the
unipolar regime ($E_{f}^{(r2)}>0$), plateaus of the ensemble average $%
\langle T\rangle $ are only partly formed when $\Delta =2t$, and the height
may not be accurately of the expected values given by
\begin{equation}
\langle T\rangle =2\times \mbox{min}(|\nu _{1}|,|\nu _{2}|)
\label{eq:tfnn_mean}
\end{equation}%
The transmission functions show large ensemble variance where $\langle
T\rangle $ has a large deviation from Eq. \eqref{eq:tfnn_mean}. Decreased
disorder strength in the junction interface, in contrast, leads to
better-developed plateaus of $\langle T\rangle $ in the unipolar regime, at
the cost of losing the quantized values of $\langle T\rangle $ and Var($T$)
in the bipolar regime. This is presented as the dash lines in Fig. \ref%
{fig:diff_Ef_r2} with $\Delta =0.5t$.

The plateaus described by Eq. \eqref{eq:tfnp_mean} in the bipolar regime and
described by Eq. \eqref{eq:tfnn_mean} in the unipolar regime are the
signature of the quantum Hall effect in a single graphene junction,\cite%
{Williams07sci} though in some cases the accuracy of the plateau is poor in
the experimental data,\cite{Williams07sci, Ozyilmaz07prl} compared with the
expected value. By taking into account disorder in the junction interface
area and a self-averaging process in the measurement, our calculations
clearly produce these plateaus. In addition, the lack of the accuracy of the
plateau height is attached to the strength of the disorder, and is reflected
in the variance of the transmission functions.

\begin{figure}[tbp]
\centering \includegraphics[width=0.4\textwidth,
height=7cm]{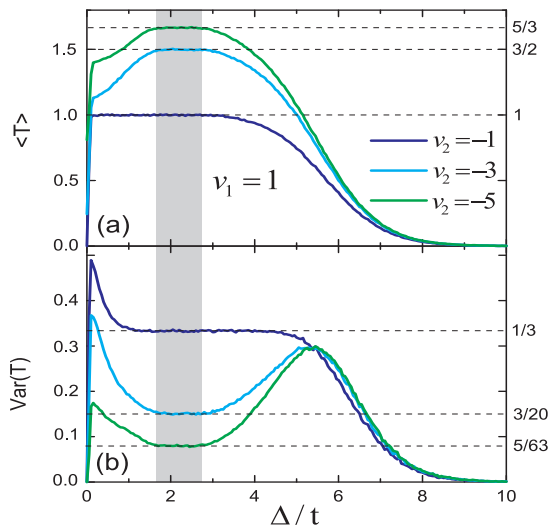} \caption{(Color online) (a) The mean
and (b) the variance of the transmission function $T$ as functions
of the disorder strength $\Delta$ in bipolar junctions. Region 1
is of n-type with the filling factor $\protect\nu_1=1$, and region
2 is of p-type with the filling factor $\protect\nu_2=-1,-3,-5$,
respectively. The shadow highlights the regime where both $\langle
T \rangle$ and Var($T$) show plateaus predicted by Eq.
\eqref{eq:tfnp_mean} and Eq. \eqref{eq:tfnp_var}, respectively.}
\label{fig:diff_disorder_np}
\end{figure}

The experimentally observed conductance plateaus of a bipolar graphene
junction in the quantum Hall regime have been explained as the result of the
complete mixing of quantum Hall edge states at the junction interface due to
scattering, and the departures of the experimental data from Eq. %
\eqref{eq:tfnp_mean} have been attributed to the incomplete mixing of edge
states.\cite{Williams07sci, Abanin07sci} We emphasize here that because the
spin-flip process is negligible in this system, the mode mixing can only
happen among states of the same spin quantum number. Thus to correctly
express $\langle T\rangle $ and Var($T$) in terms of the filling factors $%
\nu _{1}$ and $\nu _{2}$, the filling factors must be spinless, with the
spin degree of freedom included as an extra multiplier to $T$. Such
expressions, compared with the expressions using the spinful filling factors
$\pm 2$, $\pm 6$, $\pm 10$...,\cite{Abanin07sci} show no quantitative
difference as for $\langle T\rangle $, but significant differences as for
Var($T$).

The disorder dependence of the transmission functions with specific
combinations of filling factors in the bipolar regime is shown in Fig. \ref%
{fig:diff_disorder_np}. The shadowed region (roughly $1.7t<\Delta<2.8t$) is
where complete mode mixing happens in all three cases, i.e. $\nu_1=1$ and $%
\nu_2=-1,-3,-5$, respectively. The averaged transmission functions develop
into plateaus of height described by Eq. \eqref{eq:tfnp_mean} simultaneously
in this region, and the ensemble variances of the transmission functions
also develop into plateaus predicted by Eq. \eqref{eq:tfnp_var}. In the
language of the RMT,\cite{BarangerJalabert94, Beenakker97rmp} the ensembles
of scattering matrices $S$ (of a specific spin quantum number) under these
circumstances are the circular unitary ensembles (CUEs), that is, $S$
matrices in these ensembles are uniformly distributed over the unitary group
$\mathcal{U}(\nu_{\mbox{\scriptsize
total}})$. Average over the CUE is equal to an integration over the unitary
group. And it is this integration that lead to the "universal" value of the
averaged transmission function Eq. \eqref{eq:tfnp_mean}, and the universal
conductance fluctuation (UCF) given by Eq. \eqref{eq:tfnp_var}. The graphene
bipolar junctions in this disorder regime (shadowed) are nearly ideal
realizations of the quantum chaotic cavities characterized by Eq. %
\eqref{eq:tfnp_mean} and \eqref{eq:tfnp_var}.

The ensemble of $S$ matrices, however, is actually dependent on the disorder
strength $\Delta$, which represents how much the scattering potential can be
varied from one sample to another. In Fig. \ref{fig:diff_disorder_np} we see
that to the left of the shadowed region (roughly $\Delta<1.7t$), both $%
\langle T \rangle$ and Var($T$) deviate from Eq. \eqref{eq:tfnp_mean} and %
\eqref{eq:tfnp_var} with decreased $\Delta$, indicating the deviation of the
actual ensembles of $S$ matrices from the CUEs. In other words, these are
the cases of incomplete mode mixing. Notably, when $\nu_1=-\nu_2=1$ there is
an extended range of $\Delta$ where the plateau of $\langle T \rangle$ is
preserved. This is easily understood as there are only two modes to be mixed
in this case. When the disorder at the junction interface is larger than the
range of the shadowed region in Fig. \ref{fig:diff_disorder_np}, on the
other hand, the conducting edge modes in either region 1 or region 2 are
expelled from the junction interface area gradually, scattering between
modes from one side of the junction to the other is weak, both $\langle T
\rangle$ and Var($T$) decrease until the whole junction is "cut off". This
is the case of what we see in the large-$\Delta$ regime of Fig. \ref%
{fig:diff_disorder_np}.

\begin{figure}[tbp]
\centering \includegraphics[width=0.38\textwidth]{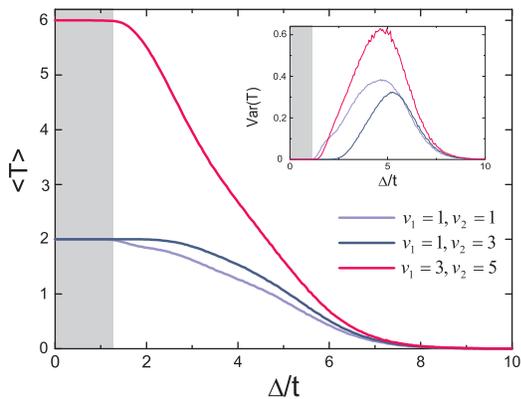}
\caption{(Color online) The averaged transmission function
$\langle T \rangle$ as a function of the disorder strength
$\Delta$ in unipolar (n-n) junctions with different combinations
of filling factors. The inset shows the corresponding variance of
$T$. The regime is shadowed where all plateaus of $\langle T
\rangle$ are preserved.} \label{fig:diff_disorder_nn}
\end{figure}

Compared with the bipolar junctions, transport in a unipolar junction with
disordered junction interface has a more straightforward picture. The
conductance of a unipolar junction is mainly contributed from the compatible
edge modes in the two regions, and the tunneling of carriers from one region
to another mainly happens near the lateral edges. The disorder at the
interface impedes the coupling of these modes to reduce the transmission
through the junction, which is contrary to the bipolar cases. The calculated
$\langle T \rangle$ and Var($T$) in the unipolar regime are shown in Fig. %
\ref{fig:diff_disorder_nn} with different combinations of filling factors.
In each case $\langle T\rangle$ exhibits a quantized plateau when the
disorder is weak (shadowed), and starts to deviate from the plateau at a
critical value of the disorder strength where Var($T$) also begins to
deviate from zero. We notice that the shadowed regions in Fig. \ref%
{fig:diff_disorder_np} and \ref{fig:diff_disorder_nn} do not overlap over
the range of $\Delta$, which implies that $\langle T\rangle$ do not develop
into plateaus simultaneously in the bipolar regime and the unipolar regime,
especially when the filling factors are large. This fact highlights the
opposite roles the disorder plays in the formation of conductance plateaus
in the bipolar and the unipolar junctions. It stabilizes the plateaus in the
bipolar case when its intensity is in a certain nonzero range, while it
tends to destroy the plateaus in the unipolar case at the same time, though
in the limit of strong disorder electrons will be blocked from tunneling
through the junction in both cases.

\begin{figure}[tbp]
\centering
\includegraphics[width=0.38\textwidth,
height=5.4cm]{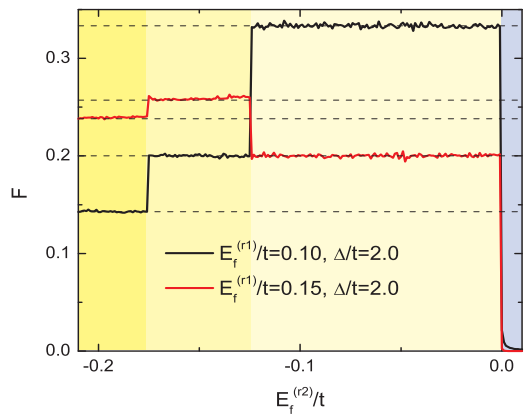} \caption{(Color online) The Fano
factors in the bipolar regime, as functions of the relative Fermi
energy $E_f^{(r2)}$, with $\protect\nu_1$ = 1 (black) and 3
(red/dark gray), respectively, and $\Delta=2t$. Plateaus predicted
by Eq. \eqref{eq:ff} are indicated by short dash lines.}
\label{fig:fanofactors}
\end{figure}

The character of the quantized transport in these junctions can be
experimentally identified by measuring the electron shot noise.\cite%
{Blanter00} The Fano factor, defined as the ratio of the actual shot noise
to the Poisson noise, is extracted from our simulation using the equation $%
F=\langle \sum T_{n}(1-T_{n})\rangle /\langle \sum T_{n}\rangle $, where the
summations are taken over different transmission eigenvalues indexed by $n$.
It is found that the Fano factors corresponding to the conductance plateaus
in the unipolar regime are identically zero, as expected from the
dissipationless transport via the quantum Hall edge modes. In the bipolar
regime, the Fano factors develop into plateaus described by
\begin{equation}
F=\frac{|\nu _{1}\nu _{2}|}{\left( |\nu _{1}|+|\nu _{2}|\right) ^{2}-1}
\label{eq:ff}
\end{equation}%
corresponding to the plateaus of $\langle T\rangle $, as shown in Fig. \ref%
{fig:fanofactors}. Eq. \eqref{eq:ff} is again a straightforward outcome of
the RMT applied to a quantum chaotic cavity with few transmission modes.\cite%
{Savin06} It is quantitatively different from the Fano factors discussed in
reference \cite{Abanin07sci}, especially when the number of the transmission
modes is small. And this is a key to examine experimentally the existence of
the UCF revealed in our simulation.

\section{Discussions and conclusion}

Before ending this paper we address the issues on the sources and
types of disorder in graphene which is relevant to this study. The
experimentally observed conductance plateaus were explained as a
result of the full mixing of these quantum Hall edge
modes.\cite{Williams07sci, Abanin07sci} The details of the mode
mixing are determined by the form of the disorder and in turn
provide information about the disorder therein. The quantum Hall
edge modes in a graphene strip can be indexed by two quantum
numbers, spin and valley isospin,\cite{Akhmerov07prl} besides the
Landau levels they belong to. Thus the mechanism of the mode
mixing inevitably involves the presence of spin-flip scattering
and/or intervalley scattering. Considering the negligible magnetic
impurities and spin-related interaction in the current graphene
samples, and ignoring the magnetic edge states reported in some
specific graphene ribbons,\cite{Son06nat}, we assume the absence
of spin-flip scattering in this study, and use the "spinless"
filling factors $\nu _{1}$ and $\nu _{2}$ ($=\pm 1,\pm 3,\pm
5\cdots $) in our expressions.\cite{note2} The presence of
intervalley scattering among the quantum Hall edge modes in the
graphene p-n junctions, however, is still a question. Previous
works on the disorder effects in graphene p-n junctions generally
assumed weak intervalley scattering, and discussed the
valley-isospin dependence of the conductance.\cite{GPNJtheories}
For example, in the work by Tworzyd{\l}o \textit{et
al.},\cite{Tworzydlo07prb} it was assumed that each impurity has
the Gaussian potential profile $U_{i}\exp
(-|\bm{r}-\bm{R}_{i}|^{2}/2\xi ^{2})$ of range $\xi $ and random
height $U_{i}\in (-\delta ,\delta )$. A large $\xi $ implies a
long range of the disorder potential which suppresses the
intervalley scattering, while a small $\xi$ implies
$\delta$-function-like disorder potential which provokes the
intervalley scattering. The former case was studied but the
plateaus observed in the experiments were not properly recovered.
That is the reason why we choose the latter case as our starting
point in this work, though the origin of the short-range disorder
potential is not completely understood. It turns out our
calculations do produce the reported conductance plateaus as in
Eq. \eqref{eq:tfnp_mean}, as well as quantities like the Fano
factors as in Eq. \eqref{eq:ff} that can be further examined
experimentally. Still it is interesting to investigate the
crossover effects between long-range disorder potential and
short-range disorder potential in the junctions, and the results
would possibly explain the not-fully-developed conductance
plateaus in the experiments.\cite{Williams07sci, Ozyilmaz07prl}
This investigation will be a subject of our later work.
Furthermore, there also exist other forms of disorder that cannot
be described by random local potential,\cite{Neto07091163,
Pereira08prb}. For example, effective gauge fields can be induced
by ripples, topological lattice defects, strains, or curvatures
\textit{etc.} in the graphene sample, and will affect the Landau
levels, especially the lowest one.\cite{Guinea08prb} We believe
that these various forms of disorder will affect the behavior of
the edge-mode mixing in various manners, but this is out of the
scope of the current work.

In short, our numerical simulation of the quantum Hall effect in graphene
p-n junctions has reproduced the quantized conductance plateaus observed in
the experiment. The UCF and quantized values of the Fano factors are found
to be accompanying the conductance plateaus in the bipolar regime, which is
well explained by the RMT of quantum transport. The bipolar graphene
junction in the quantum Hall regime mimics an ideal quantum chaotic cavity,
which is another example of the extraordinary transport character of
graphene.

\begin{acknowledgments}
This work was supported by the Research Grant Council of Hong Kong
under Grant No.: HKU 7041/07P. After we completed the present work
we were aware of a similar work by W. Long \textit{et
al.}.\cite{Long08xxx}
\end{acknowledgments}

\section{Appendix: Derivation of Eqs. (\ref{eq:tfnp_mean}), (\ref{eq:tfnp_var}), and (\ref{eq:ff})}

In this appendix, we derive Eqs. \eqref{eq:tfnp_mean},
\eqref{eq:tfnp_var}, and \eqref{eq:ff} using the random matrix
theory.

Random matrix theory of quantum transport is based on the assumption that
the scattering matrix $S$ of a chaotic cavity is uniformly distributed over
a specific group (the so-called circular ensemble) determined by the
symmetry of the Hamiltonian. Basically there are three classes of the groups.%
\cite{Dyson1962} If time-reversal symmetry is broken ($\beta =2$),
$S$ is only constrained by unitarity, which is a result of current
conservation, thus $S$ belongs to a unitary group. If
time-reversal symmetry is preserved together with the presence of
spin-rotation symmetry ($\beta =1$), then $S$ is both unitary and
symmetric: $S=S^{T}$, where the superscript $T$ indicates the
transpose of the matrix, this leads to an orthogonal group. If
time-reversal symmetry is preserved but spin-rotation symmetry is
broken ($\beta =4$), which is the case when spin-orbit interaction
is present, then $S$ is unitary and self-dual: $S=S^{R}$, where
the superscript $R$ indicates the dual of a quaternion matrix, the
group is called symplectic.

In random matrix theory, the statistics of transport properties is obtained
from the statistics of an appropriate circular ensemble. For example, the
mean of the transmission probability $T_{nm}\equiv |S_{nm}|^{2}$ is given by
\begin{equation}
\langle T_{nm}\rangle =\int d\mu (S)S_{nm}S_{nm}^{\ast }
\end{equation}%
where $n$ and $m$ stand for transmission eigen-channels. We will omit the
mathematical details of formulating the measure $d\mu (S)$ in a group space
that $S$ belongs to, and use the following two equations as being
established:\cite{math}
\begin{equation}
\langle U_{\alpha a}U_{\beta b}^{\ast }\rangle _{\mbox{\scriptsize CUE}}=%
\frac{1}{N}\delta _{\alpha \beta }\delta _{ab}
\end{equation}%
\begin{align}
& \langle U_{\alpha a}U_{\alpha ^{\prime }a^{\prime }}U_{\beta b}^{\ast
}U_{\beta ^{\prime }b^{\prime }}^{\ast }\rangle _{\mbox{\scriptsize CUE}}
\notag \\
=& \frac{1}{N^{2}-1}(\delta _{\alpha \beta }\delta _{ab}\delta _{\alpha
^{\prime }\beta ^{\prime }}\delta _{a^{\prime }b^{\prime }}+\delta _{\alpha
\beta ^{\prime }}\delta _{ab^{\prime }}\delta _{\alpha ^{\prime }\beta
}\delta _{a^{\prime }b})  \notag \\
-& \frac{1}{N(N^{2}-1)}(\delta _{\alpha \beta }\delta _{ab^{\prime }}\delta
_{\alpha ^{\prime }\beta ^{\prime }}\delta _{a^{\prime }b}+\delta _{\alpha
\beta ^{\prime }}\delta _{ab}\delta _{\alpha ^{\prime }\beta }\delta
_{a^{\prime }b^{\prime }})
\end{align}%
where $U$ is an $N$$\times $$N$ unitary matrix belonging to the circular
unitary ensemble (CUE). Since the system we are going to discuss only
involves the CUE, we will drop this notation hereafter.

Starting from these equations, the conductance $\langle G\rangle
$, its fluctuation Var($G$), and the Fano factor $F$ for the
fully-mixed quantum hall edge transport in a graphene p-n junction
can be readily calculated. We first assume no spin-flip scattering
is allowed, so that $S$ matrix for each spin component is unitary
by itself, the case including spin-flip scattering will be
discussed in what follows. Suppose there are $2$$N_{1}$ and
$2$$N_{2}$ edge states (they are also eigen-channels away from the
junction area) at the two sides of the junction, respectively, the
full $S$ matrix is of dimension $2(N_{1}+N_{2})$, but it is
divided into two uncoupled spin subspaces in the absence of
spin-flip scattering, therefore matrices in the CUE are only of
dimension $N_{\mbox{\scriptsize total}}=N_{1}+N_{2}$ in terms of
each spin subspace. Also the strict spin degeneracy implies that
$S_{nm}^{\sigma }=S_{nm}^{-\sigma }=S_{nm}$. Regarding this we
have
\begin{eqnarray} \label{eq:G}
\langle G\rangle &=& \sum\limits_{\sigma =\pm
1}\sum\limits_{n=1}^{N_{1}}\sum\limits_{m=1}^{N_{2}}\langle
T_{nm}^{\sigma }\rangle\frac{e^{2}}{h}  \notag  \\
&=& \: 2\sum\limits_{n=1}^{N_1} \sum\limits_{m=1}^{N_2}
\langle S_{nm}S^{*}_{nm}\rangle\frac{e^{2}}{h} \nonumber\\
&=& \frac{2N_{1}N_{2}}{N_{1}+N_{2}}\frac{e^{2}}{h},
\end{eqnarray}%
\begin{align}
\mbox{Var}(G)& =(\langle G^{2}\rangle -\langle G\rangle ^{2}  \notag \\
=& \left\langle \left( \sum\limits_{\sigma =\pm
1}\sum\limits_{n=1}^{N_{1}}\sum\limits_{m=1}^{N_{2}}T_{nm}^{\sigma
}\right) ^{2}\right\rangle\left( \frac{e^{2}}{h}\right) ^{2}
-\langle G\rangle ^{2}
\notag \\
= & 4\sum\limits_{n,n^{\prime }=1}^{N_{1}}\sum\limits_{m,m^{\prime
}=1}^{N_{2}} \langle S_{nm} S_{n'm'} S_{nm}^{*}
S_{n'm'}^{*}\rangle\left( \frac{e^{2}}{h}\right)^{2}
-\langle G\rangle ^{2}\nonumber\\
=& \frac{4N_{1}^{2}N_{2}^{2}}{\left( N_{1}+N_{2}\right)
^{2}\left[( N_{1}+N_{2}) ^{2}-1\right]}\left(
\frac{e^{2}}{h}\right) ^{2}.
\end{align}%
The Fano factor $F=\langle \mbox{Trace}(tt^{\dagger }(1-tt^{\dagger
}))\rangle /\langle \mbox{Trace}(tt^{\dagger })\rangle $, where $t$ is the
transmission submatrix of $S$,\cite{lastnote} and $\mbox{Trace}(tt^{\dagger
})$ is nothing but $G/\frac{e^{2}}{h}$. Thus
\begin{align}
F=& 1-\frac{N_{1}+N_{2}}{2N_{1}N_{2}}\sum\limits_{\sigma =\pm
1}\sum\limits_{n,n^{\prime }=1}^{N_{1}}\sum\limits_{m,m^{\prime
}=1}^{N_{2}}\langle S_{nm}^{\sigma }S_{n^{\prime }m}^{\sigma \;\ast
}S_{n^{\prime }m^{\prime }}^{\sigma }S_{nm^{\prime }}^{\sigma \;\ast }\rangle
\notag \\
=& \frac{N_{1}N_{2}}{\left( N_{1}+N_{2}\right) ^{2}-1}
\end{align}%
In this way we have derived the formulae expressed in Eqs.
\eqref{eq:tfnp_mean}, \eqref{eq:tfnp_var}, and \eqref{eq:ff} in
the framework of the random matrix theory.

In contrast to the spin-conserved scattering case, the mixing
among different spin modes will lead to quantitatively different
variance and Fano factors, while leaving the conductance mean
unchanged. In the random matrix theory, this is derived as
following. Still suppose there are $2$$N_{1}$ and $2$$N_{2}$ edge
states at the two sides of the junction respectively, the full $S$
matrix is still of dimension $2(N_{1}+N_{2})$, but the
spin-off-diagonal parts are now non-zero, and the $S$ matrix is
unitary only as a whole. It is straightforward to show that in
this case the formulae corresponding of Eqs. \eqref{eq:tfnp_mean},
\eqref{eq:tfnp_var}, and \eqref{eq:ff} are given by
\begin{eqnarray}
\langle G\rangle  &=&\frac{4N_{1}N_{2}}{2N_{1}+2N_{2}}\frac{e^{2}}{h}  \notag
\\
&=&\frac{2N_{1}N_{2}}{N_{1}+N_{2}}\frac{e^{2}}{h},
\end{eqnarray}%
\begin{eqnarray}
\mbox{Var}(G) &=&\frac{(4N_{1}N_{2})^{2}}{\left( 2N_{1}+2N_{2}\right)
^{2}[(2N_{1}+2N_{2})^{2}-1]}\left( \frac{e^{2}}{h}\right) ^{2}  \notag \\
&=&\frac{4N_{1}^{2}N_{2}^{2}}{\left( N_{1}+N_{2}\right)
^{2}[(N_{1}+N_{2})^{2}-1/4]}\left( \frac{e^{2}}{h}\right) ^{2},
\end{eqnarray}%
and%
\begin{eqnarray}
F &=&\frac{4N_{1}N_{2}}{\left( 2N_{1}+2N_{2}\right) ^{2}-1}  \notag \\
&=&\frac{N_{1}N_{2}}{\left( N_{1}+N_{2}\right) ^{2}-1/4},
\end{eqnarray}%
respectively. Except for a coincident equivalence of the mean
values of the conductance $\langle G\rangle $, the quantitative
differences of the UCF $\mbox{Var}(G)$ and the Fano factors $F$
between two cases, namely whether scattering is spin-conserved or
not, are obvious, especially when the values of $N_{1}$ and
$N_{2}$ are small.


\begin{thebibliography}{99}
\bibitem{note2} By the random matrix theory, the calculations including or excluding
mixing of different spin modes lead to the same expression for the
averaged conductance, but different expressions for the
conductance fluctuation and the Fano factor, which is derived in
the Appendix. And this accounts for the differences between Eqs.
(5) and (7) in this work and Eqs. (4) and (10) in the work by
Abanin and Levitov.\cite{Abanin07sci}
\bibitem{lastnote} In general $S$ can be written in terms of its transmission
and reflection parts as
\begin{equation*}
S=\left(
\begin{array}{cc}
r & t^{\prime } \\
t & r^{\prime }%
\end{array}%
\right) .
\end{equation*}
\end{thebibliography}
\end{document}